\documentclass[aip,reprint]{revtex4-1}

\usepackage{graphicx}% Include figure files
\usepackage[usenames,dvipsnames]{xcolor}
\usepackage{multirow}
\draft % marks overfull lines with a black rule on the right

\begin{document}

%\preprint{}

\title{Reference-free GIXRF-XRR as a methodology for independent validation of XRR on ultrathin layer stacks and a depth-dependent characterization} %Title of paper


%& \noindent\large{,$^{\ast}$\textit{$^{a}$} Blanka Detlefs,\textit{$^{b}$} Janis Eilbracht,\textit{$^{a}$} Yves Kayser,\textit{$^{a}$} Uwe M\"uhle,\textit{$^{c}$} Beatrix Pollakowski\textit{$^{a}$} and Burkhard Beckhoff\textit{$^{a}$}} \\%Author names go here instead of "Full 
\author{Philipp H\"onicke}
\email[]{philipp.hoenicke@ptb.de}
\affiliation{Physikalisch-Technische Bundesanstalt (PTB), Abbestr. 2-12, 10587 Berlin, Germany}
\author{Blanka Detlefs}
\affiliation{CEA-LETI, 17 rue des Martyrs, 38054 Grenoble, France.}
%\author{Janis Eilbracht}
\author{Yves Kayser}
\affiliation{Physikalisch-Technische Bundesanstalt (PTB), Abbestr. 2-12, 10587 Berlin, Germany}
\author{Uwe M\"uhle}
\affiliation{TU Dresden, Inst. of Material Science, Helmholtzstr. 7, 01062 Dresden, Germany.}
\author{Beatrix Pollakowski}
\author{Burkhard Beckhoff}
%\thanks{}
\affiliation{Physikalisch-Technische Bundesanstalt (PTB), Abbestr. 2-12, 10587 Berlin, Germany}


\date{\today}

\begin{abstract}
Nanolayer stacks are technologically very relevant for current and future applications in many fields of research. A non-destructive characterization of such systems is often performed using X-ray reflectometry (XRR). For complex stacks of multiple layers, low electron density contrast materials or very thin layers without any pronounced angular minima, this requires a full modeling of the XRR data. As such modeling is using the thicknesses, the densities and the roughnesses of each layer as parameters, this approach quickly results in a large number of free parameters. In consquence, cross-correlation effects or interparameter dependencies can falsify the modeling results. 
Here, we present a route for validation of such modeling results which is based on the  reference-free grazing incidence X-ray fluorescence (GIXRF) methodology. In conjunction with the radiometrically calibrated instrumentation of the Physikalisch-Technische Bundesanstalt the method allows for reference-free quantification of the elemental mass depositions. In addition, a modeling approach of reference-free GIXRF-XRR data is presented, which takes advantage of the quantifiable elemental mass depositions by distributing them depth dependently. This approach allows for a reduction of the free model parameters. Both the validation capabilities and the combined reference-free GIXRF-XRR modeling are demonstrated using several nanoscale layer stacks consisting of HfO$_2$ and Al$_2$O$_3$ layers.
\end{abstract}


\maketitle
 %\maketitle must follow title, authors, abstract and \pacs

\section{Introduction}
Nanoscale material systems are a relevant topic in many fields of current materials research, especially in nanoelectronics\cite{S.W.King2013,Clark2014} and energy storage applications\cite{J.Azadmanjiri2014}. Driven by the search for novel material functionalities and improved performance, the variety of investigated material combinations with respect to their elemental and structural composition is steadily growing\cite{Bunday2016,Bunday2017}. Also, the desired single layer thicknesses are in the low nanometer regime, which results in a strong additional influence of interfacial properties between adjacent materials on the integral functionality of the system. One methodology that is widely used for the characterization of such nanomaterials is X-ray reflectometry (XRR). This technique is easily available also on laboratory tools and is often used to derive thicknesses, densities and roughnesses of nanolayer stacks.

XRR is a well-established technique for sample systems with sufficiently high electron density contrast and thicknesses larger than a few nanometers\cite{Colombi_2008,J.Wernecke2014}, where the angular oscillations in XRR provide a direct and traceable access to the thickness of the thin layer. However, for more complex stacks of multiple layers, low electron density contrast or very thin layers without any pronounced angular minima, it may not be sufficient to perform XRR only\cite{A.Haase2016} as such systems require a careful modeling of the experimental data. The modeling of XRR data is usually performed by using assumed structure and the thickness, the density and the surface roughness of each layer in the stack as the optimization parameters. Remaining discrepancies due to interfacial mixing are often taken into account by adding interfacial layers\cite{J.Tiilikainen2007b} with additional parameters. Therefore, these modeling approaches quickly rely on a large number of free modeling parameters if samples with multiple nanolayers are to be investigated. This kind of approach can easily suffer from cross-correlation effects or interparameter dependencies \cite{J.Tiilikainen2007b,D.L.Gil2012} limiting stand-alone XRR data interpretation. A validation to which extent any parameter cross-correlation effects reduce the reliability of the derived modeling results is usually missing as it is not straightforward.

This issue can be addressed with the reference-free X-ray fluorescence spectrometry methodologies\cite{B.Beckhoff2007a,Beckhoff2008} of the PTB, Germany's national metrology institute. By relying on radiometrically calibrated instrumentation\cite{B.Beckhoff2009c} and reliable knowledge of the atomic fundamental parameters, no certified reference material or calibration standards are needed for a quantitative analysis of the mass deposition of an element of interest. In fact, reference-free X-ray fluorescence spectrometry can even be used to qualify reference materials or calibration samples \cite{Hoenicke2018}. As the mass depositions are the product of each materials density and thickness, they can be used to independently validate any XRR modeling result. In addition, reference-free grazing incidence X-ray fluorescence (GIXRF) \cite{M.Mueller2014} is capable to provide also depth dependent information about the sample structure \cite{P.Hoenicke2009}.

GIXRF is based on the angular and depth dependent changes of the intensity distribution within the X-ray standing wave (XSW) field arising from the interference of incident and reflected X-rays on a flat surface or interface. Due to the complementary nature of the analytical information provided by GIXRF and the dimensional information provided by X-ray reflectometry (XRR), also a combined analysis of GIXRF and XRR data is possible. This was already identified to be a promising methodological approach to reliably characterize nanostructures by DeBoer et al\cite{D.K.G.DeBoer1995} in the early 1990s.

In this work, we will demonstrate how the results of a conventional XRR modeling can be validated using the quantification capabilities of reference-free GIXRF at higher incident angles, where the XSW can be neglected \cite{M.Mueller2014}. In addition, we present an alternative modeling approach based on a hybrid reference-free GIXRF-XRR methodology. It takes advantage of the quantified elemental mass depositions for each element which can then be used as modeling constraints in order to reduce the amount of free parameters in both GIXRF and XRR evaluations. This is basically achieved by distributing the elemental mass depositions in depth into several layers, which can also intermix at interfaces.

In this work, we use the reference-free GIXRF methodology of PTB\cite{B.Beckhoff2007a,Beckhoff2008,V.Soltwisch2018} in order to gain access to the mass depositions, but of course any other first-principle or SI traceable technique which can provide this information at reasonably low uncertainties could be used. So even though we used rather sophisticated synchrotron radiation based instrumentation a very similar approach can be performed using well characterized laboratory GIXRF-XRR equipment as long as the relevant mass depositions can be derived absolutely. 

\section{Experimental}
In this work, thin Al$_2$O$_3$/HfO$_2$/Al$_2$O$_3$ layer stacks with individual thicknesses in the nanometer range have been deposited on silicon wafers with a native oxide layer.We specifically chose these two oxides as they provide a very high electron density contrast. The layers were fabricated at CEA-LETI using atomic layer deposition (ALD), which is a technique that provides very well defined and uniform layers. For both metal oxides, water vapor was used as the oxygen source during the ALD deposition process. Trimethylaluminium and HfCl$_4$ were used as precursors and the processing temperature during ALD deposition was 300 $^\circ$C. In addition to varying individual layer thicknesses, also one sample with an opposite layer sequence as well as one sample with three repetitions of the single three layer sequence was deposited. In table \ref{samples} on overview of the used samples can be found.

\begin{table}[htbp]
  \centering
  \caption{Description of the different layer sequences of the nanolaminate samples used in this work.}
    \begin{tabular}{cc}%{p{4.215em}p{14.715em}}
    \hline
    Sample & Layer sequence \\
    \hline
    S1    & Al$_2$O$_3$/  HfO$_2$ /SiO$_2$ on Si \\
    S2    & HfO$_2$/ Al$_2$O$_3$/ HfO$_2$/ SiO$_2$ on Si \\
    S3    & [Al$_2$O$_3$/ HfO$_2$]$_3$/ SiO$_2$ on Si \\
    S4    & Al$_2$O$_3$/ HfO$_2$/ Al$_2$O$_3$/ SiO$_2$ on Si \\
    S4 800 $^\circ$C   & annealed for 40 s at 800 $^\circ$C \\
    S5    & Al$_2$O$_3$/ HfO$_2$/ Al$_2$O$_3$/ SiO$_2$ on Si \\
    \hline
    \end{tabular}%
  \label{samples}%
\end{table}%

At a later stage, one pieces of the S4 wafer was thermally annealed in N$_2$ atmosphere for 40 s at 800 $^\circ$C. The annealing conditions were chosen to be identical to the work of Lan et al.\cite{X.Lan2013} in order to obtain comparable results.

\subsection{XRR characterization}

\begin{figure}
\includegraphics[width=7.5cm]{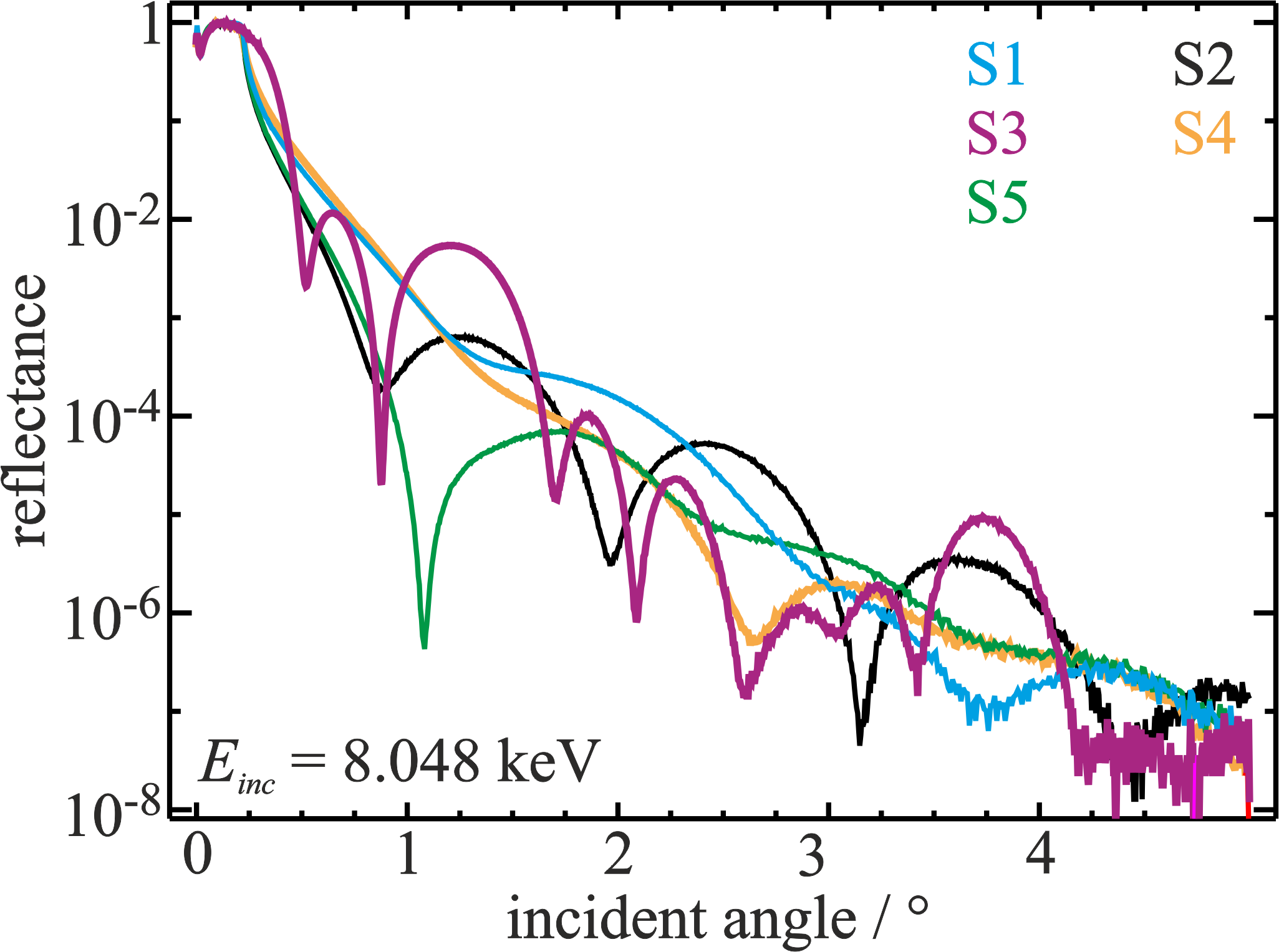}%
\caption{\label{LETI-XRR}Comparison of the measured XRR using Cu-K$\alpha$ radiation on the different samples.}%
\end{figure}

Directy after deposition, each sample was characterized at LETI by means of XRR measurements. These experiments were performed using a Bruker D8 Fabline instrument handling 300mm wafers. A Cu-K$\alpha$ X-ray source was used for this laboratory XRR characerization. The data was modeled using the Bruker LEPTOS software \cite{Ulyanenkov2004} and GenX \cite{M.Bjoerck2007}, an XRR analysis code, based on the differential evolution algorithm. Both modelings were performed by using the densities, the thicknesses and the roughnesses of each layer as the modeling parameters. In addition, also the roughness of the silicon substrate was varied. In fig. \ref{LETI-XRR}, the different XRR curves for the various samples are shown. The modeling results from the LETI XRR experiments are shown in table \ref{LETI-Fitresults}.

\subsection{GIXRF-XRR characterization}
The reference-free GIXRF-XRR experiments were carried out in the PTB laboratory at the electron storage ring BESSY II, employing the plane grating monochromator (PGM) beamline for undulator radiation\cite{F.Senf1998} as well as the four-crystal monochromator (FCM) beamline for bending magnet radiation\cite{Krumrey1998}.  At both beamlines, PTB's in-house built instrumentation\cite{J.Lubeck2013} for reference-free XRF and XRR experiments was used. The setup is installed in an ultra-high vacuum (UHV) chamber equipped with a 9-axis manipulator, allowing for a very precise sample alignment with respect to all relevant degrees of freedom. The emitted fluorescence radiation is detected by means of a calibrated\cite{F.Scholze2001a,F.Scholze2009} silicon drift detector (SDD) mounted at 90$^\circ$ with respect to the incident beam. Additional calibrated\cite{B.Beckhoff2009c} photodiodes on a separate 2$\theta$ axis allow for XRR measurements simultaneously with the reference-free GIXRF measurements as well as for the determination of the incident photon flux.

To optimize the excitation conditions for the Al-K and the fluorescence lines originating from the Hf-L3 shell, the nanolaminate samples were measured at two incident photon energies $E_{inc}$ (1.622 keV and 10.0 keV). The excitation energy of 10 keV presents the advantage that only the L3 shell of Hf can be ionized and that Coster-Kronig transitions do not occur. The excitation energy of 1.622 keV (which is below the Si-K absorption edge) results in a drastically reduced spectral background for the Al-K line and in the suppression of any secondary excitation channels. Thus, the selection of these excitation conditions allows for the lowest achievable uncertainties in the Al and Hf quantification. For each photon energy, both a reference-free GIXRF and a XRR measurement were conducted in parallel.

\section{Results and Discussion}

\subsection{Validation of XRR modeling results}

\begin{table}[htbp]
  \centering
  \caption{Overview on the various parameters as determined from a modeling of the XRR data shown in fig. \ref{LETI-XRR}. The results of sample S3 are not shown.}
    \begin{tabular}{|c|c|c|c|c|c|c|}
    \hline
          &       & \textbf{S1} & \textbf{S2} & \textbf{S4} & \textbf{S5} & Unit \\
    \hline
    \multirow{4}[8]{*}{\textbf{1st Layer}} & \textbf{Material} & Al$_2$O$_3$ & HfO$_2$  & Al$_2$O$_3$ & Al$_2$O$_3$ &  \\
\cline{2-7}          & \textbf{Thickness} & 1.82 & 0.65 & 1.29  & 2.48 &  nm \\
\cline{2-7}          & \textbf{Density} & 2.75 & 7.53 & 3.18  & 2.88 &  gcm$^{-3}$ \\
\cline{2-7}          & \textbf{Roughness} & 0.3   & 0.29  & 0.46  & 0.46  &  nm \\
    \hline
    \multirow{4}[8]{*}{\textbf{2nd Layer}} & \textbf{Material} & HfO$_2$  & Al$_2$O$_3$ & HfO$_2$  & HfO$_2$  &  \\
\cline{2-7}          & \textbf{Thickness} & 1.28 & 2.78 & 1.64 & 0.7   &  nm \\
\cline{2-7}          & \textbf{Density} & 8.99 & 2.84 & 9.24 & 7.40 &  gcm$^{-3}$ \\
\cline{2-7}          & \textbf{Roughness} & 0.21  & 0.28  & 0.18  & 0.2 & nm \\
    \hline
    \multirow{4}[8]{*}{\textbf{3rd Layer}} & \textbf{Material} & ---   & HfO$_2$  & Al$_2$O$_3$ & Al$_2$O$_3$ &  \\
\cline{2-7}          & \textbf{Thickness} & ---   & 0.74 & 1.30 & 2.34 &  nm \\
\cline{2-7}          & \textbf{Density} & ---   & 6.98 & 3.57 & 2.97 &  gcm$^{-3}$ \\
\cline{2-7}          & \textbf{Roughness} & ---   & 0.23  & 0.28  & 0.39  &  nm \\
    \hline
    \multirow{3}[6]{*}{\textbf{SiO$_2$}} & \textbf{Thickness} & 1.10 & 0.86 & 1.32 & 1.13 &  nm \\
\cline{2-7}          & \textbf{Density} & 2.02 & 1.86 & 2.20 & 2.08 &  gcm$^{-3}$ \\
\cline{2-7}          & \textbf{Roughness} & 0.39  & 0.41  & 0.61  & 0.49  &  nm \\
    \hline
    \textbf{Substrate} & \textbf{Roughness} & 0.2   & 0.27  & 0.2   & 0.15  &  nm \\
    \hline
    \end{tabular}%
  \label{LETI-Fitresults}%
\end{table}%

In table \ref{LETI-Fitresults}, the obtained modeling results from the LETI XRR experiments are shown. As already mentioned, the densities, the thicknesses and the roughnesses of each layer as well as the substrates roughness served as the modeling parameters. The XRR results are in line with expectations, e.g. that the derived material densities are somewhat lower than the corresponding bulk densities \cite{S.Sintonen2014} and that the roughnesses are in the order of half a nanometer. In addition, also the achieved agreement between the experimental and the modeled XRR curves (not shown here) is very good and does not indicate any issues. However, as there are at least ten independent modeling parameters and the features in the experimental XRR data are not always very pronounced, one may expect parameter correlation effects. The remaining question is now how one can evaluate how severe they influence the results.

\begin{figure*}
\includegraphics[width=15cm]{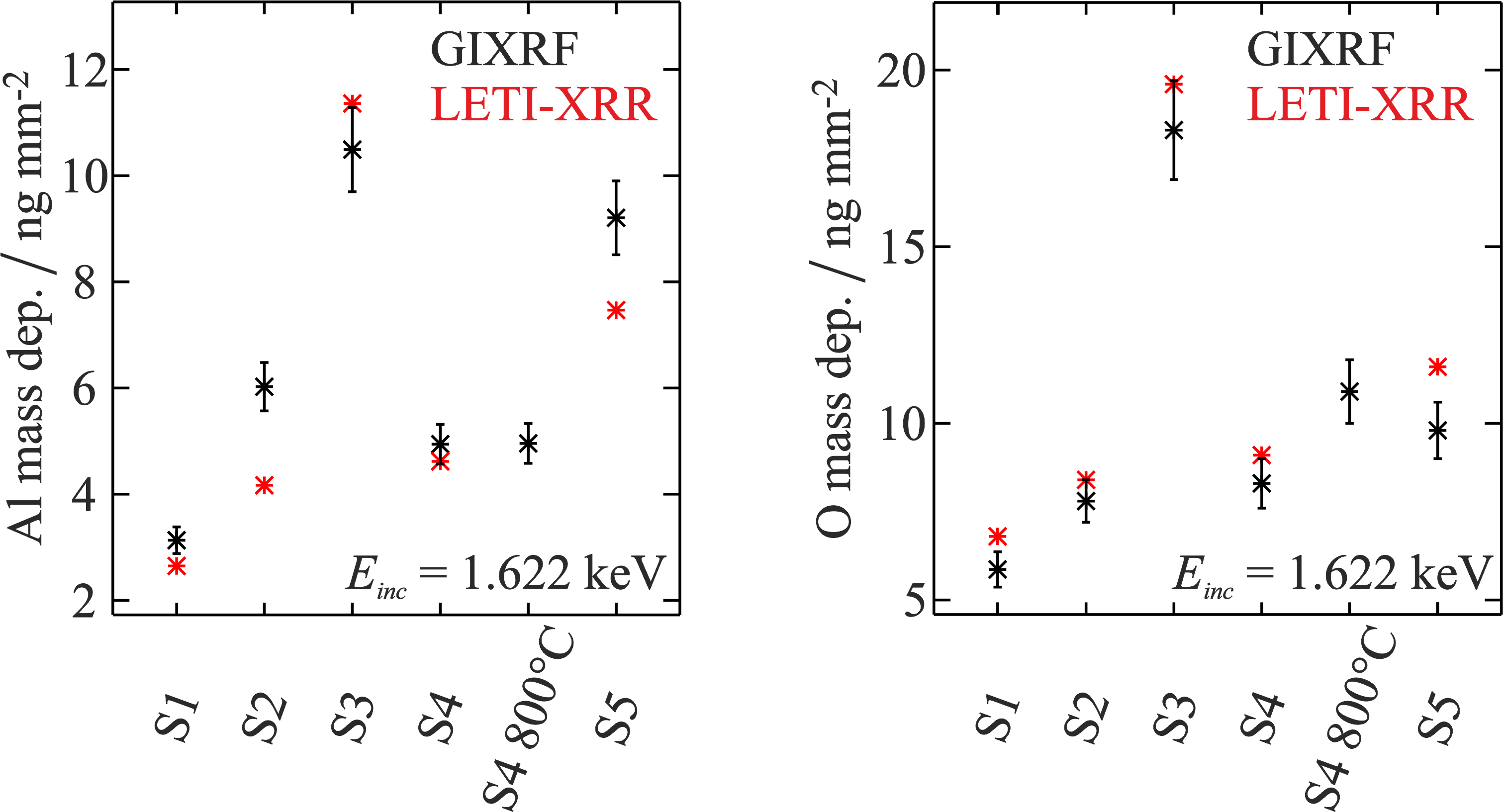}%
\caption{\label{Quant_Al_O}Comparison of quantified elemental mass depositions for Al (left) and O (right) versus the calculated data using the modeling results of the LETI-XRR data. The reference-free GIXRF quantification was performed for incident angles above 4$^\circ$.}%
\end{figure*}

\begin{figure}
\includegraphics[width=7.5cm]{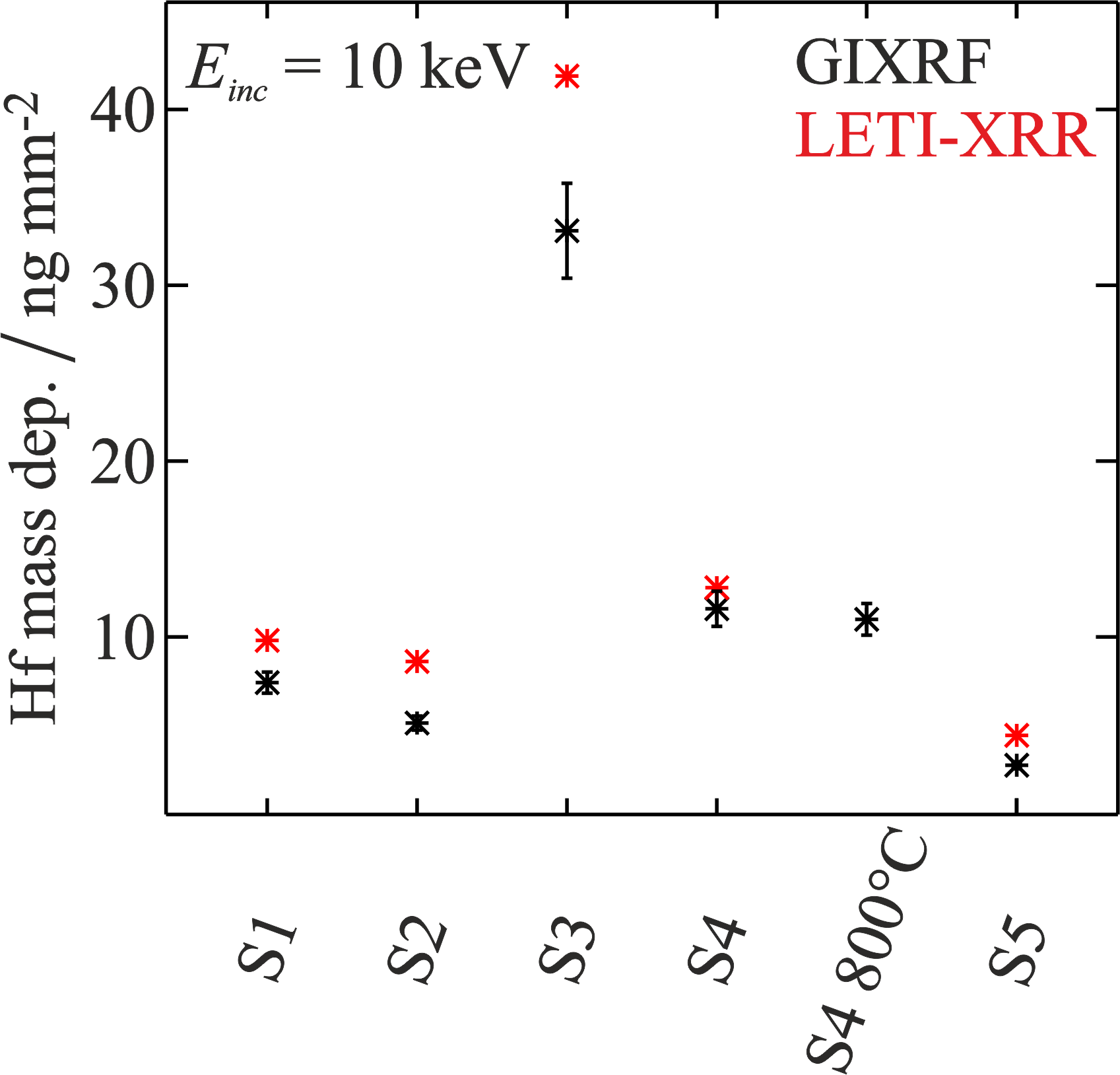}%
\caption{\label{Quant_Hf}Comparison of quantified elemental mass depositions for Hf versus the calculated data using the modeling results of the LETI-XRR data. The reference-free GIXRF quantification was performed for incident angles above 4$^\circ$.}%
\end{figure}

One way to perform a validation of such modeling results is to calculate each materials mass deposition ($=$ product of material density and thickness) from the thicknesses and densities as determined by the XRR data evaluation. These mass depositions can then be compared to mass depositions obtained from e.g. quantitative GIXRF experiments. In the following, we have calculated the corresponding total elemental mass depositions of Al, Hf and O from each resulting parameter set by multiplying the respective modeled densities and thicknesses assuming stoichiometric materials\cite{K.Kukli2002} and then use reference-free GIXRF experiments for an independent validation.
At incident angles far above the critical angle for total external reflection the fluorescence intensity modulations due to the interplay of the XSW field and the nanolayer stack vanish and a direct quantification of the mass depositions can be performed without any structural modeling. To do so in reference-free GIXRF experiments, the recorded fluorescence spectra are deconvolved using the known detector response functions\cite{F.Scholze2009} for the relevant fluorescence lines as well as for the background contributions. A direct and traceable quantification of the mass depositions can then be performed as presented in ref. \cite{M.Mueller2014} using Sherman's equation. The necessary instrumental parameters, e.g. the solid angle of detection or the incident photon flux are known due to the usage of the well-known physically calibrated instrumentation\cite{Beckhoff2008}. The relevant fundamental parameters are partially taken from databases \cite{T.Schoonjans2011} or derived from dedicated experiments in the case of the oxygen\cite{P.Hoenicke2016a} and aluminum \cite{Beckhoff2008} K-shell as well as the Hf-L3 shell (according to ref. \cite{M.Kolbe2012}) fluorescence yields. The derived mass depositions for Al, O and Hf are shown in figs. \ref{Quant_Al_O} and \ref{Quant_Hf} as black stars. A relative experimental uncertainty between 8 \% and 9 \%, which is dominated by the relative uncertainties of the fundamental parameters involved in the quantification is achieved. The calculated mass depositions for the XRR modeling results are also shown in figs. \ref{Quant_Al_O} and \ref{Quant_Hf} as red stars. 

The direct comparison of the XRR modeling and the reference-free GIXRF derived mass depositions reveals discrepancies for some samples. With respect to Al, the results only agree within the corresponding uncertainties for sample S4. For all other samples, there is a larger mismatch. For oxygen, only sample S5 shows a significant mismatch but on all samples, the XRR modeling results yield too much oxygen. For Hf, the agreement is similar as for Al. These deviations show, that despite the high electron density contrast between HfO$_2$ and Al$_2$O$_3$ and despite the superior deposition capabilities of ALD, the XRR modeling still results in substantial deviations.
The main reason for these differences originates in the fact that both the material densities and thicknesses are free model parameters and, thus, the materials mass deposition can be varied by the modeling. Consequently any shortcomings of the used layer model, e.g. surface contamination or interface diffusion or uncertainties of the used optical constants, e.g. due to missing fine structure close to absorption edges will be compensated to some extent by a wrong adjustment of the densities and thicknesses. If there are enough free parameters this is not easy to detect, as the overall agreement between calculated XRR and the experimental data is often very good. 

For these reasons, the need for an external validation is high especially for cases, where multiple very thin layers are to be characterized as shown here. It should also be noted that the experimental XRR data for the samples in this work shows rather prominent features (see fig. \ref{LETI-XRR}). Still, the deviations between modeled mass depositions and the real ones is relatively large as shown. So one should expect this to be even more of an issue if the XRR data has less prominent features due to even thinner materials or less electron density contrast.

\subsection{Modeling of reference-free GIXRF-XRR data}
In addition to such a validation of XRR modeling results, the reference-free GIXRF-XRR data measured at the two incident photon energies $E_{inc}$ (1.622 keV and 10.0 keV) also allows for a depth dependent combined modeling of the layer structure. Here, the total mass depositions and, thus, the products of the respective layer thicknesses and densities are known from the reference-free quantification at high incident angles as described earlier. Thus, the densities and thicknesses are not allowed to vary independently because the information on the elemental mass deposition can be used by distributing them in depth into separate layers. Each layer density is a modeling parameter and the corresponding thickness is then derived from the respective mass deposition. In addition, the optical constants of a given material also scale with its density. As a result, the number of free parameters is reduced.

\begin{figure*}
\includegraphics[width=15cm]{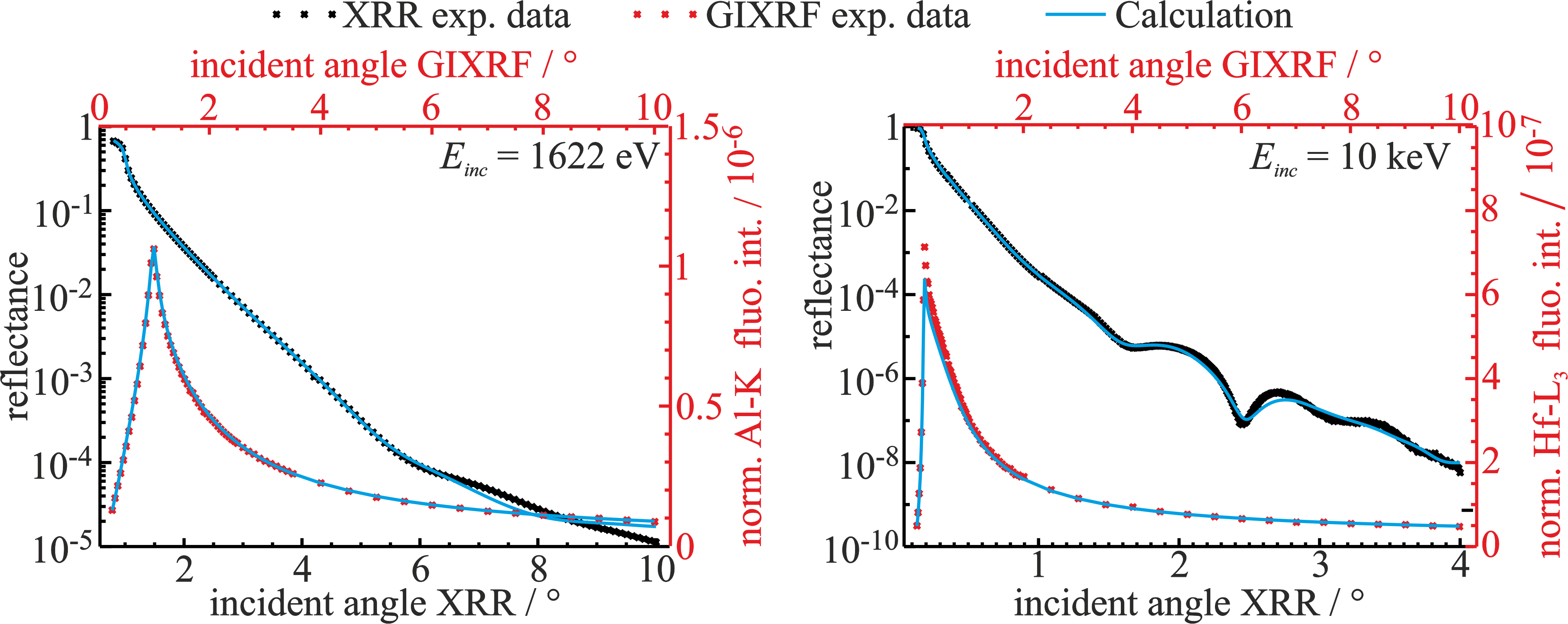}%
\caption{\label{Fit_S4_8}Comparison of the full experimental data set, consisting of an XRR (black stars and axes) and a GIXRF curve (red stars and axes) for each photon energy and the respective modeling results (blue solid lines) for sample S4 800 $^\circ$C.}%
\end{figure*}

 In addition, some corrections to take into account the respective uncertainties of the quantified mass depositions or of the used tabulated optical constants for the bulk materials are necessary. This is realized by applying scaling factors, which are allowed to vary by the respective parameters' relative uncertainty around unity. To take into account any interfacial mixing of two adjacent layers, additional intermixing coefficients for each interface are implemented. They determine the width where the materials change symmetrically from one to the other and can range from zero (no intermixing) to 1 (fully intermixed layers). 

The hybrid modeling routine for reference-free GIXRF-XRR is modeling the full data set of two XRR and two GIXRF curves at once (see fig. \ref{Fit_S4_8}) in order to take full advantage of the complementary nature of GIXRF and XRR. It first assumes density values for each layer in the stack (including a carbonaceous surface contamination layer). Using the previously quantified mass depositions, the resulting thickness values for each layer are calculated. With these layer thicknesses, concentration depth profiles for each layer are defined. Depending on the respective intermixing coefficient, these depth profiles can overlap at the interfaces. The concentration depth profiles are then used to calculate depth profiles for each optical constant ($\delta$ and $\beta$) at the two used photon energies. Here, bulk optical constants for Al$_2$O$_3$, HfO$_2$, SiO$_2$ and Si from \cite{T.Schoonjans2011} were used and also scaled with the modeled material densities. If an intermixing is present, the effective optical constants are calculated accordingly by means of a linear combination.

The full layer stack is then separated into thin sublayers in order to calculate both the resulting XRR curves for both photon energies as well as the XSW for each photon energy. A PTB in-house developed software package (XSWini\cite{Pollakowski}) was used here, as it could be directly implemented into the modeling routine. The derived intensity distribution within the XSW is then numerically integrated in conjunction with the calculated concentration depth profiles and all other relevant instrumental and fundamental parameters to calculate the angular fluorescence profiles for Al and Hf as shown in reference \cite{P.Hoenicke2009}.

\begin{figure}
\centering
  \includegraphics[width=7.2cm]{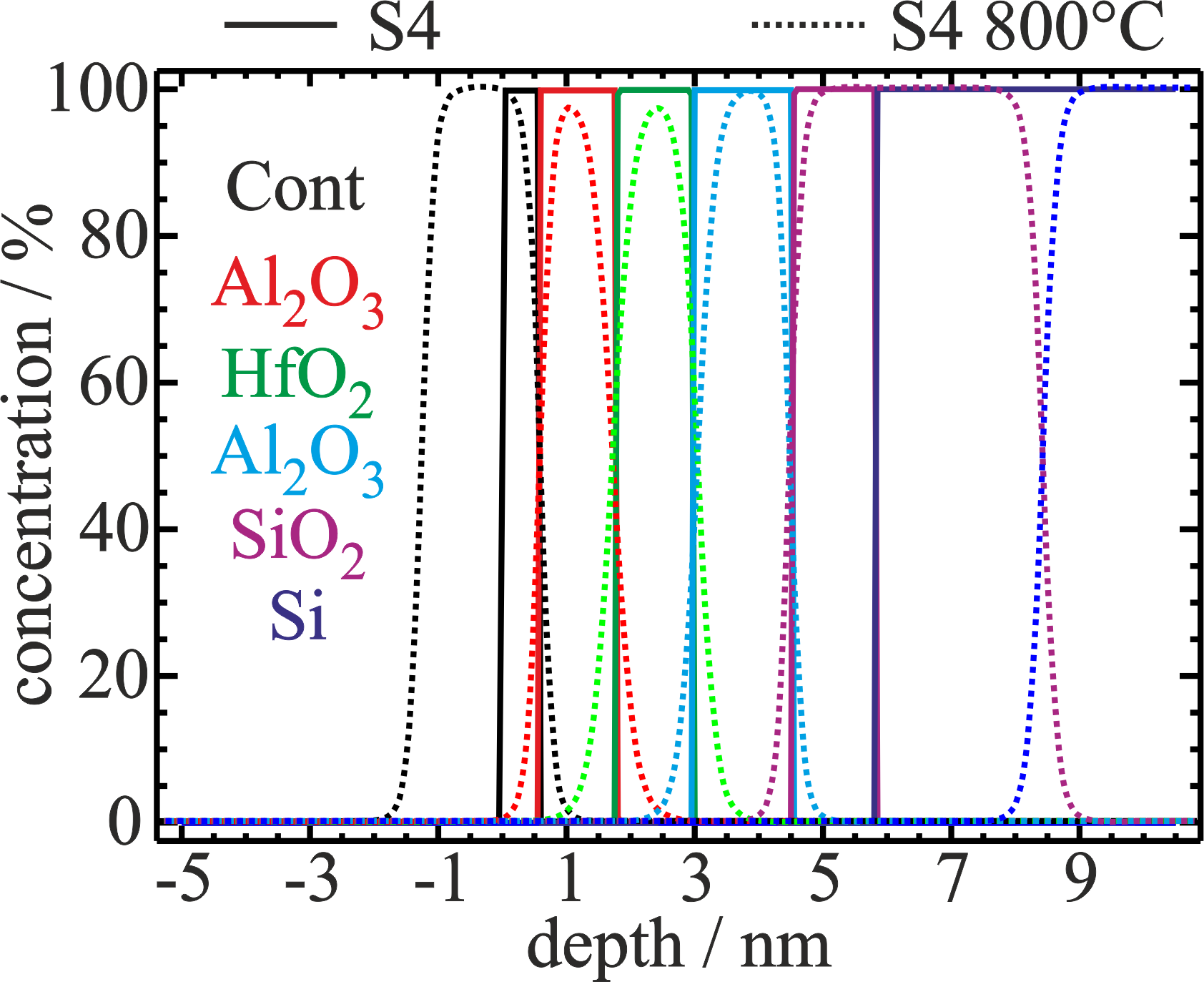}
  \caption{Comparison of the concentration depth profiles determined using the reference-free GIXRF-XRR modeling of samples S4 (solid lines) and the annealed S4 800 $^\circ$C (annealed for 40 s at 800 $^\circ$C, dotted lines).}
  \label{fgr:S4-S4-8}
\end{figure}

Using this procedure, the samples S4 and the annealed S4 800 $^\circ$C have been analyzed and the corresponding concentration depth profiles obtained are shown in fig. \ref{fgr:S4-S4-8}. The solid lines correspond to the layer stack of sample S4 whereas the dotted lines to sample S4 800$^\circ$C (annealed for 40 s at 800 $^\circ$C). An increase of the interfacial intermixing for the annealed sample is clearly visible for all interfaces. The non-annealed sample shows no relevant intermixing, which is in line with the expectations\cite{Knez2012} for such ALD depositions. The observed diffusion-driven symmetric intermixing for the annealed sample is verified by the findings in the work of Lan et al.\cite{X.Lan2013}. One should also note the increase in the modeled thickness of the SiO$_2$ layer on the annealed sample, which is in line both with the increase of the quantified oxygen mass deposition shown in fig. \ref{Quant_Al_O} and also with TEM images (not shown) of the annealing sample series. The growing oxide layer at the interface to the Si substrate is a known effect during annealing when HfO$_2$ is present and very small oxygen contaminations within the annealing atmosphere are sufficient to result in the observed SiO$_2$ growth\cite{N.Miyata2003,S.Ferrari2006}.

In summary, the combined modeling approach using each materials concentration depth profile instead of distinct layers with additional interface layers allows to derive information about the intermixing due to the thermal annealing even on such thin layers. As this would not be easily possible with GIXRF alone or with conventionally modeled XRR, the present approach provides an improved strategy for the characterization of ultrathin layers and layer stacks.

\section{Conclusions}
In summary, this work demonstrates how the widely used conventional modeling approach for XRR data obtained on very thin layers can suffer from parameter cross-correlation effects which can mask unexpected sample changes or incomplete model assumptions. This can be a crucial issue when characterizing ultra-thin layered samples and it occurs due to the resulting large number of free model parameters. As a result, these modeling strategies often provide very well reproduced experimental data but still erroneous results, which are then hard to be revealed. As many software implementations for both commercial and research tools are using this conventional approach, this issue must be addressed. Using thin nanolaminate layer stacks with Al$_2$O$_3$ and HfO$_2$ as layer materials, we have uncovered these unfavorable effects and also present both a validation scheme and a new hybrid modeling scheme. The external validation of the modeled elemental mass depositions helps to benchmark the conventional modeling results and, thus, to reveal the negative cross-correlation effects. For a more reliable modeling, the mass depositions are directly used in the presented hybrid GIXRF-XRR approach in order to reduce the risk of such hindering parameter cross-correlations.

In this respect, we derived the total elemental mass depositions using our reference-free quantification scheme\cite{B.Beckhoff2007a,Beckhoff2008,M.Mueller2014,P.Hoenicke2015} to set-up a hybrid reference-free GIXRF-XRR modeling. It uses the information about the elemental mass depositions, which can be used to reduce the degrees of freedom within the modeling. This can lead to a more reliable interpretation of the experimental data as compared to the conventional modeling approaches and compared to single XRR or GIXRF analysis. 

It also should be noted, that the presented methodology does not require synchrotron radiation sources and is transferable also to laboratory instruments. These instruments can also provide an access to the elemental mass depositions if they are well calibrated or one simply uses other quantitative methods, e.g. Rutherford backscattering spectrometry\cite{C.Jeynes2016} in order to determine the mass depositions. These can then be brought into the modeling of the laboratory GIXRF-XRR experimental data. Thus, the presented quantitative hybrid GIXRF-XRR approach combines the non-destructive and in-line capable GIXRF and XRR techniques with sufficient reliability to reveal unexpected changes to the sample structure as it has a reduced amount of degrees of freedom. 

% If in two-column mode, this environment will change to single-column format so that long equations can be displayed. 
%\begin{widetext}
%$$\mbox{put long equation here}$$
%\end{widetext}

\begin{acknowledgments}
This research was performed within the EMPIR projects Aeromet and Adlab-XMet. The financial support of the EMPIR program is gratefully acknowledged. It is jointly funded by the European Metrology Programme for Innovation and Research (EMPIR) and participating countries within the European Association of National Metrology Institutes (EURAMET) and the European Union.
\end{acknowledgments}

\bibliography{literature2}

\end{document}